\documentstyle[12pt]{article}
\begin{document}
\title{Modelling structural relaxation in polymeric glasses
using the aggregation--fragmentation concept}

\author{Aleksey D. Drozdov\\
Institute for Industrial Mathematics\\
4 Hanachtom Street\\
Beersheba, 84311 Israel}
\date{}
\maketitle

\begin{abstract}
Governing equations are derived for the kinetics of physical aging
in polymeric glasses.
An amorphous polymer is treated as an ensemble of cooperatively
rearranged regions (CRR).
Any CRR is thought of as a string of elementary clusters (EC).
Fragmentation of the string may occur at random time at any border
between ECs.
Two string can aggregate at random time to produce a new string.
The processes of aggregation and fragmentation are treated as thermally
activated, and the rate of fragmentation is assumed to grow
with temperature more rapidly than that for coalescence.
This implies that only elementary clusters are stable at
the glass transition temperature $T_{\rm g}$, whereas
below $T_{\rm g}$, CRRs containing several ECs remain stable as well.
A nonlinear differential equation is developed for the distribution
of CRRs with various numbers of ECs.
Adjustable parameters of the model are found by fitting experimental
data for polycarbonate, poly(methyl methacrylate),
polystyrene and poly(vinyl acetate).
For all materials, fair agreement is established between observations
and results of numerical simulation.
For PVAc, the relaxation spectrum found by matching data in a calorimetric
test is successfully employed to predict experimental data in
a shear relaxation test.
\end{abstract}

\section{Introduction}

This paper is concerned with the kinetics of structural relaxation
(physical aging) in amorphous glassy polymers.
Slow dynamics in out-of-equilibrium disordered media
[supercooled liquids, structural (including polymeric) glasses,
disordered ferromagnets and antiferromagnets, 
orientational glasses,
vortices in superconductors, 
dipolar glasses,
liquid crystalline colloids,
spin glasses, etc.]
has attracted substantial attention in the past decade,
see, e.g., surveys \cite{EAN96,BCK98,Kob99,HVD99} and the references
therein.
Unlike most previous studies which focused on equilibrium thermodynamics
of glass-like systems \cite{MP00},
the present work deals with the evolution of physical
properties of amorphous polymers quenched from some temperature $T_{0}$
above the glass transition temperature $T_{\rm g}$ to a temperature
$T$ in the sub--$T_{\rm g}$ region and isothermally annealed
at the temperature $T$.
Changes in the mechanical response, specific volume, configurational entropy
and other features of glassy polymers caused by thermal jumps
have been widely studied (both experimentally and theoretically)
in the past four decades.
We would like to mention here seminal works by Kovacs \cite{Kov64,KAH79}
and Struik \cite{Str78}.
For a detailed review, the reader is referred to \cite{MK89}.

Despite a number of publications concerned with the evolution
of internal structure of disordered media after thermal jumps,
see \cite{BCK98} for a survey,
it is difficult to mention a model (phenomenological or molecular)
which correctly predicts experimental data
(for example, memory and chaos phenomena in
cyclic thermal tests \cite{BCL99a,BCL99b})
and establishes an adequate correspondence between observations in various
(mechanical, dielectric, calorimetric, dilatometric, etc.)
experiments, see, e.g., \cite{DLK96,DP98,SW98,CHE98,CFH98,HSH99}
and the references therein.

An important class of constitutive models for structural relaxation
in amorphous materials employs the concept of coarsening in disordered
systems \cite{BCK98}.
First models belonging to this class (the so-called droplet
models) have been proposed a decade ago by Koper and Hilhorst \cite{KH88}
and Fisher and Huse \cite{FH88} to predict the response of spin glasses.
According to their approach, an arbitrary macro-region of a disordered
medium is treated as an ensemble of droplets (domains with non-regular
boundaries possessing non-zero fractal dimensions) where spins are aligned
with one of the ground states.
These regions alter with time because of slow coarsening,
and the characteristic size of a region grows as a power of time
\cite{KH88} or as a logarithm of time \cite{FH88}.
For a detailed review of phenomenological models for the phase-ordering
kinetics, we refer to \cite{Bra94}.

Another approach to the description of domain growth
has been developed by Ben-Naim and Krapivsky \cite{BK95,BK96,BK97,BK98}.
The present study employs basic ideas of their works to predict
enthalpy recovery in amorphous polymers and to study the effect
of waiting time $t_{\rm w}$ on the material response
in mechanical (relaxation) tests.
We would like to stress two issues that distinguish our constitutive
equations from those derived by Ben-Naim and Krapivsky:
\begin{enumerate}
\item
In the analysis of coalescence, we presume that for any domain,
the rate of aggregation with another domain is proportional to
the probability density (not to the concentration) of regions
with a certain volume (length).
This makes solutions of constitutive equations independent of the
total number of elementary clusters at the initial instant.

\item
We combine the aggregation-fragmentation theory with the concept of
internal time.
This allows fair approximation of experimental data for the relaxing
enthalpy to be established (the conventional approach fails
to adequately describe enthalpy recovery during 4 to 5 decades of time).
\end{enumerate}
The concept of internal time is widely used to describe thermo-mechanical
response of polymers (for a review, the reader is referred to \cite{Dro98}).
In this study, we employ a model with an entropy-driven material clock.
This approach was proposed by Struik \cite{Str78} and was successfully
used to predict shear-thickening of polymer solutions in \cite{Dro97}.

With reference to the theory of cooperative relaxation \cite{AG65},
an amorphous polymer is thought of as an ensemble of cooperatively
rearranging regions (CRR) that relax at random times as they
are thermally agitated.
A CRR is thought of as a globule consisting of scores of strands of
long chains \cite{Sol98}.
The characteristic length of a CRR in the vicinity of the glass transition
temperature amounts to several nanometers \cite{RN99}.

We introduce a minimal number of strands which permits
rearrangement, $\nu$, and suppose that the number of strands
in any CRR reads $n\nu$, where $n$ is an integer
and $\nu$ is the number of strands in an elementary cluster (EC).
The concept of elementary domains was first introduced by
Adam and Gibbs \cite{AG65} for amorphous polymers
and by Kadanoff \cite{Kad66} for spin glasses.
For a discussion of the Kadanoff conjecture
(near the critical point, large blocks of spins behave like individual
spins) see \cite{LD98,Bak98}.
Description of the aging process in glassy polymers in terms
of models with finite numbers of states was suggested
by Chow \cite{Cho83} and Robertson \cite{Rob78,RSC84}.

Unlike previous studies where the domain growth was interpreted
as random motion of their boundaries annihilated at the contact
points, see, e.g., \cite{BK98},
we treat structural recovery as a result of two (thermally activated)
processes with different dependences of their rates on temperature:
fragmentation of large regions and aggregation of small ones.
An indirect confirmation of this picture is provided by experimental data
in calorimetric tests at various temperatures $T_{k}$ in the
sub--$T_{\rm g}$ region.
Observations reveal that if $T_{2}>T_{1}$,
the relaxing enthalpy $\Delta H$ in a test with the temperature $T_{1}$
exceeds that in a test with the temperature $T_{2}$ at small times,
and the inverse is true at large times.

The exposition is organized as follows.
In Section 2, constitutive equations are derived for enthalpy recovery
in amorphous glassy polymers.
These equations are verified in Section 3 by comparison with experimental
data for polycarbonate, polystyrene, poly(methyl methacrylate)
and poly(vinyl acetate).
The purpose of fitting observations is two-fold: (i) to confirm that
the model can correctly predict measurements, and (ii) to analyze the
effect of temperature $T$ on material parameters.
In Section 4, the model is subject to a more elaborate evaluation:
after determining adjustable parameters by matching data in calorimetric
tests, we use the probability density of CRRs with various energies
to fit the material response in a static mechanical test.
To make this possible, a model is derived for stress relaxation in
disordered media based on the traps concept (for detail,
we refer to \cite{Dro99,Dro00}).
Some concluding remarks are formulated in Section 5.

\section{Constitutive equations for domain growth}

The following hypotheses are introduced to develop a master
equation for the concentration of CRRs:
\begin{enumerate}
\item
Structural relaxation in a disordered medium is governed by
two processes at the micro-level: fragmentation of CRRs and
their aggregation.

\item
Denote by $P(t,n)$ the number of CRRs (per unit mass) at time $t$ 
consisting of $n+1$ elementary clusters.
The quantity $P(t,n)$ is the mass concentration of CRRs containing
$n$ borders between ECs subject to fragmentation (to simplify
the analysis, any CRR is thought of as a linear ``string'' of ECs).
The function $P(t,n)$ obeys the conservation law
\begin{equation}
\sum_{n=0}^{\infty} (n+1)P(t,n)=\Xi_{0},
\end{equation}
where $\Xi_{0}$ is the number of ECs per unit mass.

\item
The fragmentation process is characterized by its rate $\gamma$
that equals the number of fragmentation acts per boundary between ECs
per unit time.
We assume $\gamma$ to be a function of the current temperature $T$ only.
The rate of changes in the quantity $P(t,n)$ induced by fragmentation
is given by
\begin{equation}
V_{\rm f}(t,n)=\gamma \Bigl [ -nP(t,n)+2\sum_{m=n+1}^{\infty} P(t,m)\Bigr ].
\end{equation}
The first term on the right-hand side of Eq. (2) determines the number of
CRRs (containing $n$ boundaries) destroyed by fragmentation, whereas the
other term is the rate of creation of CRRs with $n$ boundaries caused
by fragmentation of relaxing regions containing larger number of elementary
clusters.
The physical meaning of formula (2) is discussed in detail in \cite{Red90}.

\item
Coalescence of relaxing regions occurs when CRRs (with $n$ and $m$
borders between ECs) merge and create a new CRR (with $n+m+1$ borders).
The rate of aggregation of an individual CRR with $n$ boundaries between
ECs with CRRs with $m$ borders is proportional to the density of CRRs
with $m$ boundaries (the ratio $\varphi(t,m)$ of the number of CRRs
with $m$ boundaries to the total number of CRRs per unit mass)
and to the number of random meetings, $\Gamma_{n,m}$, for two CRRs
with $n$ and $m$ boundaries.
The rate of changes in the quantity $P(t,n)$ induced by coalescence
reads
\begin{eqnarray}
V_{\rm c}(t,n) &=&
-P(t,n) \Bigl [ \sum_{m=0,m\neq n}^{\infty} \Gamma_{n,m} \varphi(t,m)
+2\Gamma_{n,n}\varphi(t,n)\Bigr ]
\nonumber\\
&& +\sum_{m=0}^{n-1} \Gamma_{m,n-m-1}P(t,m)\varphi(t,n-m-1)
\end{eqnarray}
with
\begin{equation}
\varphi(t,n)=\frac{P(t,n)}{\sum_{m=0}^{\infty} P(t,m)}.
\end{equation}
The first two terms on the right-hand side of Eq. (3) determine the number
of CRRs destroyed by aggregation with other CRRs (the coefficient 2 in
the second term means that two relaxing regions with $n$ borders
disappear when they meet one another).
The last term in Eq. (3) characterizes the rate of creation of CRRs with
$n$ boundaries by aggregation of two relaxing regions with smaller numbers
of ECs.

\item
The rate $\Gamma_{n,m}$ by which regions with $n$ and $m$ boundaries
aggregate decays exponentially with the growth of the indices $n$ and $m$,
\begin{equation}
\Gamma_{n,m}=L\gamma\exp \Bigl [-\lambda (n+m)\Bigr ],
\end{equation}
where $L$ and $\lambda$ are material parameters.
Assumption (5) is conventional for the treatment of (short-range)
interactions between CRRs, but it differs from standard
relations based on the theory of anomalous diffusion,
see, e.g., \cite{KHG95} and the references therein, which lead to
the power law dependence of $\Gamma_{m,n}$ on $n$ and $m$.
Our numerical analysis shows that replacement of Eq. (5) by a power
law does not affect significantly the accuracy of fitting.

\item
We employ the concept of material (reduced) time and suppose that $\gamma$
is the rate with respect to some internal time $\tau$.
The rate of fragmentation with respect to the ``universal'' time $t$ reads
\begin{equation}
\gamma_{0}=a\gamma,
\end{equation}
where $a$ is a shift factor.

\item
For polymers with an entropy-driven material clock \cite{Str78}, we set
\begin{equation}
\ln a=-\kappa_{0} s,
\end{equation}
where $\kappa_{0}$ is some material parameter and $s$ is
the configurational entropy per EC.

\item
The configurational entropy $s$ is defined by the Boltzmann's formula
\begin{equation}
s(t)=-k_{B}\sum_{n=0}^{\infty} \varphi(t,n)\ln \varphi(t,n),
\end{equation}
where $k_{B}$ is Boltzmann's constant.
It is worth noting some difference between Eq. (8)
and conventional relations, see, e.g., \cite{AM88}.
In Eq. (8) the configurational entropy $s$ is defined per elementary
cluster, whereas traditional formulas define configurational entropy
per CRR.
\end{enumerate}

It follows from Eqs. (2) to (7) that the current concentration of CRRs
(per unit mass) is governed by the differential equation
\begin{eqnarray}
\frac{\partial P}{\partial t}(t,n)
&=& \gamma_{0} \exp \Bigl ( -\kappa s(t)\Bigr )
\biggl \{ -nP(t,n)+2\sum_{m=n+1}^{\infty} P(t,m)
-\frac{L}{\sum_{m=0}^{\infty} P(t,m)}
\nonumber\\
&& \times \biggl [ P(t,n) \sum_{m=0,m\neq n}^{\infty} P(t,m)
\exp\Bigl (-\lambda (n+m)\Bigr ) +2 P^{2}(t,n)\exp \Bigl (-2\lambda n\Bigr )
\nonumber\\
&& -\exp \Bigl ( -\lambda (n-1)\Bigr )\sum_{m=0}^{n-1} P(t,m)P(t,n-m-1)
\biggr ] \biggr \},
\end{eqnarray}
where $\kappa=k_{B}\kappa_{0}$.

The study is confined to one-step ``quench-and-wait'' tests,
when a polymer equilibrated at some temperature $T_{0}>T_{\rm g}$
is quenched to a temperature $T<T_{\rm g}$ and is annealed at
the temperature $T$,
\begin{equation}
T(t)=\left \{\begin{array}{ll}
T_{0}, & t<0,\\
T, & t>0.
\end{array}\right .
\end{equation}
Assuming that above the glass transition temperature $T_{\rm g}$,
only elementary clusters may exist, we postulate
\begin{equation}
P(0,n)=\Xi_{0}\delta_{n,0},
\end{equation}
where $\delta_{n,m}$ is the Kronecker delta.
The solution $P(t,n)$ of Eqs. (8), (9) and (11) is independent of the
total number of ECs, $\Xi_{0}$, and it is determined by 4 adjustable
parameters: $L$, $\gamma_{0}$, $\kappa$ and $\lambda$.

Let $h(t)$ be the configurational enthalpy per EC.
Assuming $h$ to be expressed in terms of the configurational entropy $s$
by the conventional relationship,
\[ \frac{\partial h}{\partial s}=T, \]
and integrating this equality for the thermal program (10),
we find that
\begin{equation}
h(t)=Ts(t) \qquad
(t>0).
\end{equation}
Multiplying Eq. (12) by the concentration of ECs, $\Xi_{0}$,
we calculate the configurational enthalpy per unit mass.
Assuming the relaxing enthalpy $\Delta H$ (measured in calorimetric tests)
to coincide with the configurational enthalpy, we arrive at the formula
\begin{equation}
\Delta H(t)=-\Lambda \sum_{n=0}^{\infty} \varphi(t,n)\ln \varphi(t,n)
\end{equation}
with
\begin{equation}
\Lambda=k_{B}T\Xi_{0}.
\end{equation}

\section{Comparison with observations in calorimetric tests}

We begin with experimental data for polycarbonate.
For a detailed description of samples and the experimental procedure,
see \cite{BCB87}.
Adjustable parameters are found by matching observations
using the steepest-descent procedure.
Figure~1 demonstrates that the model correctly predicts experimental
data at two temperatures in the sub--$T_{\rm g}$ region.
Setting $T=T_{\rm g}=420$~K in Eq. (14),
using mass density $\rho=1.196$ g/cm$^{3}$ \cite{HKJ90},
and the value $\Lambda=0.8$ J/g found by fitting observations,
we obtain
\begin{equation}
\Xi_{0}=1.15\cdot 10^{26}
\quad
\mbox{m}^{-3}.
\end{equation}
This value is in accord with $\Xi_{0}=3.6\cdot 10^{26}$ m$^{-3}$
for polytetrafluoroethylene \cite{DSF98},
but it is less than concentrations of holes measured for a series of
polycarbonates using positron lifetime spectroscopy \cite{BWK99}
(for comparison, we assume that any EC may be associated with a micro-hole).
To explain this discrepancy, we recall that (i) Bohlen et al.
\cite{BWK99} presumed a Gaussian distribution of holes with a peak
far above the zero volume, whereas our calculations demonstrate
that the probability density of CRRs substantially differs from
the Gaussian ansatz (see Figure~2),
and (ii) the value (15) is for temperatures in the vicinity of $T_{\rm g}$,
while the PALS measurements were carried out at room temperature.

Suppose that the effect of temperature on the rate of fragmentation
$\gamma_{0}$ is described by the Arrhenius formula
\begin{equation}
\gamma_{0}(T)=\gamma_{\ast}\exp \Bigl (-\frac{\Delta E}{RT}\Bigr ),
\end{equation}
where $\Delta E$ is some activation energy and $R$ is the universal
gas constant.
Applying Eq. (16) to two temperatures, $T_{1}$ and $T_{2}$, in the
sub--$T_{\rm g}$ region, we find that
\begin{equation}
\Delta E=\frac{RT_{\rm g}^{2}}{T_{2}-T_{1}}
\ln\frac{\gamma_{0}(T_{2})}{\gamma_{0}(T_{1})}.
\end{equation}
Taking values of $\gamma_{0}(T_{k})$ determined by fitting experimental
data (Figure~1), we obtain $\Delta E=174.1$ kcal/mol, which is
in fair agreement with $\Delta E=173.6$ kcal/mol
found by shift of relaxation curves measured at various
temperatures in the vicinity of the glass transition temperature
\cite{Dro98}.

According to Figure~1, an increase in temperature $T$ leads to a sharp
decrease in the relative rate of aggregation $L$ and to an increase in
the parameter $\lambda$.
This conclusion is in fair agreement with our hypothesis that
above the glass transition temperature $T_{\rm g}$ only elementary
clusters exist (because the rate of merging vanishes, the growth of
CRRs caused by their coalescence is prohibited).
It is worth noting that an increase in $\lambda$ with temperature
observed in experiments makes questionable models of aggregation
that are based on the diffusion mechanism, see \cite{KHG95},
provided that the coefficient $\lambda$ in Eq. (5) may be
associated (at least, to some extent) with the decrease in
the diffusion coefficient induced by the growth of mass of domains
to be aggregated.

The equilibrium distributions depicted in Figure~2 are characterized
by their mean value $M_{1}$ and variance $M_{2}$,
\begin{equation}
M_{1}=\sum_{n=0}^{\infty} n\varphi(\infty,n),
\qquad
M_{2}=\sum_{n=0}^{\infty} (n-M_{1})^{2}\varphi(\infty,n).
\end{equation}
Numerical simulation demonstrates that the quantities $M_{1}$
and $M_{2}$ substantially increase with a decrease in $T$.
This is in accord with the conventional approach to the description
of structural relaxation \cite{BCK98,MP00}
which presumes a substantial growth in the roughness of the energy
landscape with a decrease in temperature
(which, in turn, results in a monotonic increase in the variance
of the distribution of CRRs with $\Delta T=T_{\rm g}-T$).

We proceed with matching experimental data for polystyrene using
the above algorithm.
For a description of the experimental procedure, we refer to \cite{RM83}.
Figure~3 reveals fair agreement between observations at two temperatures
in the sub--$T_{\rm g}$ region and results of numerical analysis.
Equation (14) with $T=T_{\rm g}=373.0$~K, $\rho=1.04$ g/cm$^{3}$
and $\Lambda=1.1$ J/g results in $\Xi_{0}=2.55\cdot 10^{26}$ m$^{-3}$,
which is in acceptable agreement with PALS measurements.
It follows from Eq. (17) and Figure~3 that the activation energy
$\Delta E=136.4$ kcal/mol, which is rather close to the value
$\Delta E=170.5$ kcal/mol found by shift of creep curves
in the sub--$T_{\rm g}$ region \cite{SZ80}.
The equilibrium distributions of CRRs containing various
numbers of elementary clusters are plotted in Figure~4
which demonstrate that an increase in $\Delta T$ leads to an
increase in $M_{1}$ and $M_{2}$.
This is in qualitative agreement with conventional models for
aging that predict the growth of roughness of the energy landscape
with a decrease in temperature.

Experimental data for polycarbonate and polystyrene (only at two
temperatures in the sub--$T_{\rm g}$ region)
are not sufficient to analyze the effect of temperature
on the kinetics of structural relaxation.
To study changes in adjustable parameters caused by the annealing
temperature $T$, we match observations for poly(methyl methacrylate).
For a detailed description of samples and the experimental
procedure, we refer to \cite{CF93}.
We begin approximation with measurements at the lowest temperature
$T=375.0$~K and determine parameters $L$, $\gamma_{0}$, $\kappa$,
$\lambda$ and $\Lambda$ using a version of the steepest-descent
algorithm.
Afterwards, we fix $\Lambda$ (an analog of the number
of elementary clusters per unit mass) and fit data at other temperatures
using 4 constants: $L$, $\gamma_{0}$, $\kappa$ and $\lambda$.
The assumption that the parameter $\kappa$ is temperature-dependent
differs the treatment of data for poly(methyl methacrylate) from that
for polycarbonate and polystyrene (for which
$\kappa$ was assumed to be independent of temperature).
This is done to ensure better quality of matching experimental data.
Figures~5 to 10 demonstrate fair agreement between observations and
results of numerical simulation.

The rate of fragmentation $\gamma_{0}$ is plotted versus the increment
of temperature $\Delta T$ in Figure~11.
This figure shows that experimental data are correctly approximated
by the ``linear'' function
\begin{equation}
\log \gamma_{0}=a_{0}-a_{1}\Delta T
\end{equation}
with adjustable parameters $a_{k}$.
Equations (16) and (19) imply that in the vicinity of the glass
transition temperature $T_{\rm g}$, the activation energy $\Delta E$
is given by
\[ \Delta E=a_{1}RT_{\rm g}^{2}\ln 10. \]
It follows from this formula and Figure~11 that $\Delta E=70.81$
kcal/mol, which is in excellent agreement with data for activation
energy obtained by shift of creep and relaxation curves at various
temperatures.
For example, using WLF parameters for poly(methyl methacrylate)
\cite{AMS72},
we arrive at the value $\Delta E=83.2$ kcal/mol,
which is close to our result.
This is in contrast with the Cowie--Ferguson model \cite{CF93}
which results in the value $\Delta E=164.8$ kcal/mol for the same set
of experimental data.

The parameter $L$ (which characterizes the ratio of the rate of
aggregation to the rate of fragmentation) is depicted in Figure~11
versus the increment of temperature $\Delta T$.
Experimental data are fairly well approximated by the linear function
\begin{equation}
L=b_{0}+b_{1}\Delta T
\end{equation}
with adjustable parameters $b_{k}$.
Equation (20) implies that $L$ monotonically increases with a decrease
in temperature $T$, which means that the rate of aggregation of elementary
clusters grows with a departure from the glass transition temperature.

The dimensionless parameters $\lambda$ and $\kappa$ are plotted
versus the increment of temperature $\Delta T$ in Figure~12.
Experimental data are approximated by the functions
\begin{equation}
\lambda=c_{0}-c_{1}\Delta T,
\qquad
\log \kappa=C_{0}-C_{1}\Delta T,
\end{equation}
where $c_{k}$ and $C_{k}$ are adjustable parameters.
Figure~12 demonstrates that the quantities $\lambda$ and $\kappa$
increase with temperature.
Far below the glass transition point (about 30~K below $T_{\rm g}$)
the parameter $\lambda$ vanishes, which results in an essential
simplification of Eq. (9) (whose coefficients become independent
of $n$ and $m$, which allows a method of generating
functions to be applied to develop analytical solutions \cite{KB00}).
Explicit solutions to the governing equations are, however, beyond
the scope of the present study.

It is worth noting that phenomenological models for
physical aging in polymers presume that the effects of temperature
and structure may be separated, see, e.g., \cite{KAH79}.
Results of numerical simulation reveal that this hypothesis is valid
when the increment of temperature $\Delta T$ is not too small
(say, it exceeds 15~K) which implies that $\kappa$ becomes weakly
dependent on temperature.

Conventional constitutive equations for structural relaxation in
amorphous polymers employ the random energy model \cite{RB90},
according to which the distribution of relaxing regions with various
energies is Gaussian with temperature-dependent mean value
and standard deviation.
Fitting experimental data obtained in mechanical tests
for a number of glassy polymers show that the standard
deviation of energy of CRRs, $\Sigma$,
linearly decreases with temperature and vanishes at some critical
temperature $T_{\rm cr}$ slightly above $T_{\rm g}$ \cite{Dro00}.
Because the present study is based on the assumption that
only elementary clusters are stable at the glass transition temperature
(which implies that $T_{\rm cr}=T_{\rm g}$),
our purpose now is to demonstrate that
\[ \Sigma=\sqrt{M_{2}} \]
linearly increases with $\Delta T$ in the sub--$T_{\rm g}$ region.
Figure~13 shows that with an acceptable level of accuracy the
graphs $M_{1}(\Delta T)$ and $\Sigma(\Delta T)$ are linear.
Approximation of experimental data by the functions
\begin{equation}
M_{1}=m_{1}\Delta T,
\qquad
\Sigma=m_{2}\Delta T
\end{equation}
yields $m_{2}=0.15$~K$^{-1}$.
This result is in good agreement with $m_{2}=0.09$~K$^{-1}$
for polycarbonate and $m_{2}=0.53$~K$^{-1}$ for poly(vinyl acetate)
found by fitting observations in static and dynamic mechanical
tests \cite{Dro00}.

Hitherto, only average characteristics (first moments and entropy)
of the distribution of CRRs have been compared with observations.
To reveal that the distribution of relaxing regions $\varphi(t,n)$
itself is in agreement with measurements,
we analyze experimental data for poly(vinyl acetate) in calorimetric
and mechanical (relaxation) tests.
For a description of the experimental procedure, see \cite{CFH98}.

We begin with matching observations in a calorimetric test
at $T=303$~K.
Figure~14 demonstrates fair correspondence between experimental data
and results of numerical simulation.
Setting $T=T_{\rm g}=315$~K in Eq. (14)
and using the values $\rho=1.182$ g/cm$^{3}$ \cite{KVR98}
and $\Lambda=1.6$ J/g (Figure~14), we arrive at
$\Xi_{0}=3.24\cdot 10^{26}$ m$^{-3}$,
which is in good accord with the concentration of holes measured by PALS
for polytetrafluoroethylene \cite{DSF98}.
All adjustable parameters for PVAc (except for the rate of fragmentation
$\gamma_{0}$) are similar to those for PC, PS and PMMA.
The quantity $\gamma_{0}$ is smaller than that for other polymers.
This may be explained by the differences in the glass transition
temperatures of materials under consideration
(glass transition points for PC,PS and PMMA substantially exceed
$T_{\rm g}$ for PVAc)
and the fact that fragmentation and aggregation of CRRs are thermally
activated.

To fit data in mechanical tests, stress--strain relations
should be derived for amorphous glassy polymers under short-term
mechanical loading (compared to the waiting time $t_{\rm w}$).
In this work, we briefly sketch the development of constitutive equations
confining ourselves to uniaxial relaxation tests.
Greater detail of derivations can be found in \cite{Dro00}.

\section{The mechanical response in relaxation tests}

In accord with the hoping concept, see, e.g., \cite{Sol98},
a relaxing region is treated as a point
located at the bottom level of its potential well on the energy
landscape.
Rearrangement of CRRs is modeled as hops of relaxing regions
to higher energy levels.
Hops occur at random times, and they are driven by thermal fluctuations.
Below the glass transition temperature, energy barriers between
potential wells are assumed to be so high that
thermally agitated CRRs cannot leave their traps.
Adopting the transition-state theory \cite{Gol69},
we suppose that rearrangements occur when CRRs reach
some liquid-like (reference) energy level.
The position of this level is not fixed, but it slowly ascends
in time approaching some limiting value as $t_{\rm w}\to \infty$.

Any potential well is described by its depth $w>0$
with respect to the position of the reference energy level
at the initial instant (immediately after the quench).
Following our treatment of CRRs as aggregates composed
of integer numbers of elementary clusters, we assume that
the set of available energies $w$ is countable,
\[ w=w_{1},w_{2},\ldots,w_{n},\ldots, \]
and the value $w_{n}$ is proportional to the volume (length)
of an appropriate CRR (string),
\begin{equation}
w_{n}=\alpha n,
\end{equation}
where $\alpha>0$ is an adjustable parameter.

Let $q(\omega)d\omega$ be the probability for a CRR
to reach (in a hop) the energy level that exceeds its
bottom level by an energy $\omega^{\prime}$ located in the interval
$[ \omega,\omega+d\omega ]$.
Referring to the extreme value statistics \cite{BCK98},
we set
\[ q(\omega)=A\exp(-A\omega), \]
where $A$ is a material constant.
The probability to reach the reference state in a hop for a CRR trapped
in a potential well with the depth $w$ reads
\[ Q(t_{\rm w},w) =\int_{w+\Omega(t_{\rm w})}^{\infty} q(\omega)d\omega
=\exp \Bigl [-A \Bigl (w+\Omega(t_{\rm w})\Bigr )\Bigr ] . \]
Here $\Omega(t_{\rm w})$ is the increment of the reference energy level
after the waiting time $t_{\rm w}$ with respect to that at the initial
instant, and we take into account that the duration of conventional
mechanical tests is negligible compared to $t_{\rm w}$.

The rate of hops in a potential well, $\Gamma$, is defined
as the number of hops (of an arbitrary intensity) per unit time.
Assuming $\Gamma$ to be independent of $w$ (according to the theory of
thermally activated processes, $\Gamma$ is a function
of the current temperature $T$ only),
and multiplying the rate of hops $\Gamma$ by the probability of reaching
the liquid-like state in a hop $Q(t_{\rm w},w)$, we arrive at the
Eyring formula for the rate of rearrangement \cite{Eyr36},
\begin{equation}
R(t_{\rm w}, w)=\Gamma_{0}(t_{\rm w}) \exp (-A w),
\qquad
\Gamma_{0}=\Gamma\exp \Bigl (A\Omega(t_{\rm w})\Bigr ).
\end{equation}
The rate of rearrangement, $R(t_{\rm w},w)$, can be thought of as
the ratio of the number of rearranging regions (per unit time) to the number
of CRRs to be rearranged,
\begin{equation}
R(t_{\rm w}, w)=-\frac{1}{\Xi(t,t_{\rm w}, w)}
\frac{\partial \Xi}{\partial t}(t,t_{\rm w},w).
\end{equation}
For relaxation tests, the quantity $\Xi(t,t_{\rm w},w)$ is the number
of CRRs (per unit mass) located in cages with energy $w$ that have not
been rearranged until time $t$ (which is measured from the beginning
of the test).
Integration of Eq. (25) implies that
\begin{equation}
\Xi(t,t_{\rm w}, w) =\Xi(0,t_{\rm w}, w) \exp
\Bigl [ -R(t_{\rm w},w) t\Bigr ].
\end{equation}
The initial condition for Eq. (26) is given by
\begin{equation}
\Xi(0,t_{\rm w}, w_{n})=N(t_{\rm w})\varphi(t_{\rm e},n),
\end{equation}
where
\[ N(t_{\rm w})=\sum_{n=0}^{\infty} P(t_{\rm w},n) \]
is the number of CRRs (per unit mass) after annealing for the
time $t_{\rm w}$.

The natural configuration of a CRR after rearrangement is assumed
to coincide with the deformed configuration of the viscoelastic medium
at the instant of rearrangement.
This implies that rearranged CRRs are stress-free in a relaxation test,
and only non-rearranged regions have non-zero mechanical energies.
The strain energy density of an amorphous polymer (per unit mass)
under uniaxial loading, $U$, equals the sum of mechanical energies
of individual regions.
A CRR is thought of as a linear elastic medium with the potential
energy of deformation
\[ \frac{1}{2} \mu\epsilon^{2}, \]
where $\mu$ is the rigidity and $\epsilon$ is the macro-strain
(for definiteness, shear deformation is considered).
Combining this expression with Eqs. (23), (24), (26) and (27),
we find that
\begin{eqnarray}
U(t,t_{\rm w}) &=&
\frac{1}{2}\mu \epsilon^{2} \sum_{n=1}^{\infty} \Xi(t,t_{\rm w}, w_{n})
\nonumber\\
&=& \frac{1}{2}\mu N(t_{\rm w})\epsilon^{2}
\sum_{n=1}^{\infty} \varphi(t_{\rm w},n)\exp \Bigl [-\Gamma_{0}(t_{\rm w})
\exp (-\alpha_{0}n)t \Bigr ],
\end{eqnarray}
where $\alpha_{0}=\alpha A$.
The stress $\sigma$ is expressed in terms of the mechanical energy $U$
by the formula
\[ \sigma=\rho \frac{\partial U}{\partial \epsilon}. \]
This equality together with Eq. (28) implies that
\begin{equation}
G(t,t_{\rm w})=G_{0}(t_{\rm w})\sum_{n=1}^{\infty}
\varphi(t_{\rm w},n)\exp \Bigl [-\Gamma_{0}(t_{\rm w})
\exp (-\alpha_{0}n)t\Bigr ],
\end{equation}
where $G(t,t_{\rm w})=\sigma(t)/\epsilon$ is the current shear modulus,
$G_{0}(t_{\rm w})=\mu N(t_{\rm w})$ is the initial modulus.

To verify the model, we calculate the distribution of CRRs,
$\varphi(t_{\rm w},n)$, using Eqs. (8), (9) and (11) with
adjustable parameters found by fitting observations
in the calorimetric test, and match experimental data in the
relaxation test by Eq. (29).
Given a constant $G_{0}$, the material parameters $\Gamma_{0}$
and $\alpha_{0}$ are calculated using the steepest-descent procedure.
The initial shear modulus $G_{0}$ is determined by the least-square
algorithm.
Figure~15 demonstrates fair agreement between observations and
results of numerical simulation.

It is worth noting that our constitutive equations employ two independent
time-scales for the description of structural recovery in polymeric glasses.
One scale is entirely determined by the distribution function
$\varphi(t_{\rm w},n)$.
The other time-scale characterizes the ascent of the liquid-like energy
level with respect to the energy landscape, and it is portrayed by
the function $\Omega(t_{\rm w})$.
This approach is in agreement with that recently proposed by Tanaka
\cite{Tan99}, where two order parameters were used to predict slow
dynamics in supercooled liquids.

\section{Conclusions}

A model is derived for structural relaxation in amorphous glassy polymers
after thermal jumps.
A polymeric glass is treated as an ensemble of cooperatively
rearranging regions whose concentration changes with time because
of their fragmentation and aggregation.
A CRR is modeled as a string (linear chain) of elementary clusters.
Fragmentation of the string may occur at random time at any border
between elementary clusters with equal probability.
Aggregation of relaxing regions occurs at random time as well.
The rate of coalescence for two CRRs decreases exponentially
with the growth of their sizes.
With a decrease in temperature $T$, the rates of fragmentation and
aggregation decrease, but the rate of fragmentation reduces more
rapidly.
This implies that only elementary clusters exist at the glass
transition temperature $T_{\rm g}$, whereas in the sub--$T_{\rm g}$
region, CRRs consisting of several ECs may be stable as well.

To verify constitutive equations, we fit experimental data for
relaxing enthalpy for polycarbonate, polystyrene, poly(methyl methacrylate)
and poly(vinyl acetate).
Fair agreement is demonstrated between observations
in calorimetric tests and results of numerical analysis.
Material parameters found by fitting measurements are in good
accord with those determined experimentally in other tests.

To establish correspondence between observations in calorimetric
and mechanical tests, we find material parameters for PVAc
by fitting relaxing enthalpy, determine the relaxation spectrum
and use this spectrum to match data in mechanical (static) test.
An acceptable agreement between experimental data for the shear modulus
and numerical predictions confirms our believe that the model may be
employed for the analysis of physical aging in tests,
where several experimental methods are applied simultaneously.

\section*{Acknowledgement}

The work was supported by the Israeli Ministry of Science through
grant 1202--1--98.

\newpage
\section*{List of figures}
\noindent
Figure~1: The relaxation enthalpy $\Delta H$ J/g versus time $t$ h
for polycarbonate annealed at the temperature $T$~K.
Circles: experimental data \protect{\cite{BCB87}}.
Solid lines: numerical simulation with $\Lambda=0.8$ J/g and $\kappa=1.0$.
Curve~1: $T=408.0$, $L=80.0$, $\lambda=0.3$, $\gamma_{0}=10^{3}$ h$^{-1}$;
curve~2: $T=413.0$, $L=5.0$, $\lambda=1.2$, $\gamma_{0}=1.2\cdot 10^{4}$
h$^{-1}$
\vspace*{2 mm}\\
\noindent
Figure~2: The equilibrium distribution of CRRs
$\varphi=\varphi(\infty,n)$ for polycarbonate.
Unfilled circles: $T=408$~K, $M_{1}=4.9328$, $M_{2}=9.8743$.
Filled circles: $T=413$~K, $M_{1}=0.9157$, $M_{2}=0.6758$
\vspace*{2 mm}\\
\noindent
Figure~3: The relaxation enthalpy $\Delta H$ J/g versus time $t$ h
for polystyrene annealed at the temperature $T$~K.
Circles: experimental data \protect{\cite{RM83}}.
Solid lines: numerical simulation with $\Lambda=1.1$ J/g and $\kappa=1.0$.
Curve~1: $T=363.0$, $L=80.0$, $\lambda=0.3$, $\gamma_{0}=10^{3}$ h$^{-1}$;
curve~2: $T=366.0$, $L=35.0$, $\lambda=0.5$, $\gamma_{0}=4.4\cdot 10^{3}$
h$^{-1}$
\vspace*{2 mm}\\
\noindent
Figure~4: The equilibrium distribution of CRRs
$\varphi=\varphi(\infty,n)$ for polystyrene.
Unfilled circles: $T=363$~K, $M_{1}=4.9328$, $M_{2}=9.8743$.
Filled circles: $T=366$~K, $M_{1}=2.7906$, $M_{2}=3.6336$
The relaxation enthalpy $\Delta H$ J/g versus time $t$ h
for poly(methyl methacrylate) at $T=375.0$~K
(unfilled circles: experimental data \protect{\cite{CF93}};
solid line: numerical simulation)
and the final distribution of CRRs $\varphi=\varphi(t_{\rm w},n)$
with $t_{\rm w}=10^{2.5}$ h
\vspace*{2 mm}\\
\noindent
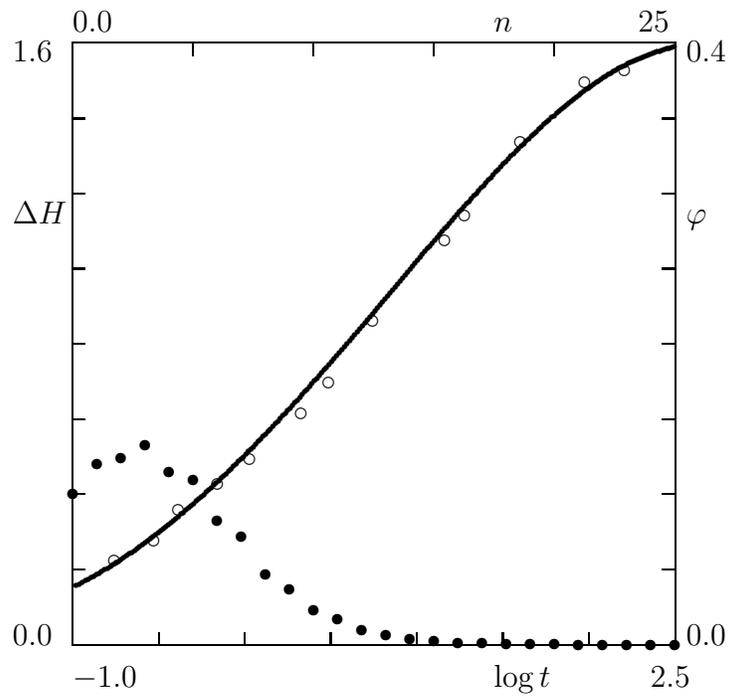
Figure~5: The relaxation enthalpy $\Delta H$ J/g versus time $t$ h
for poly(methyl methacrylate) at $T=375.0$~K
(unfilled circles: experimental data \protect{\cite{CF93}};
solid line: numerical simulation)
and the final distribution of CRRs $\varphi=\varphi(t_{\rm w},n)$
with $t_{\rm w}=10^{2.5}$ h
\vspace*{2 mm}\\
\noindent
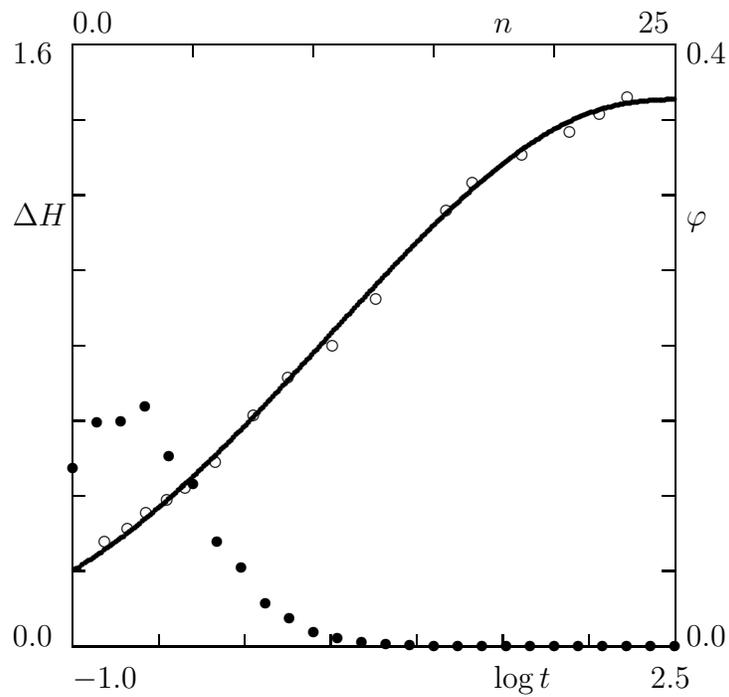
Figure~6:The relaxation enthalpy $\Delta H$ J/g versus time $t$ h
for poly(methyl methacrylate) at $T=377.5$~K
(unfilled circles: experimental data \protect{\cite{CF93}};
solid line: numerical simulation)
and the final distribution of CRRs $\varphi=\varphi(t_{\rm w},n)$
with $t_{\rm w}=10^{2.5}$ h
\vspace*{2 mm}\\
\noindent
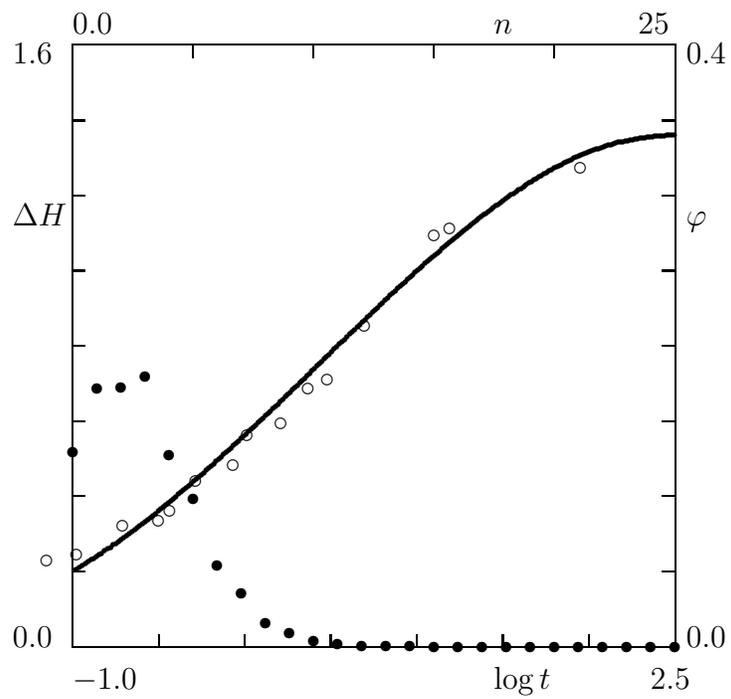
Figure~7: The relaxation enthalpy $\Delta H$ J/g versus time $t$ h
for poly(methyl methacrylate) at $T=380.0$~K
(unfilled circles: experimental data \protect{\cite{CF93}};
solid line: numerical simulation)
and the final distribution of CRRs $\varphi=\varphi(t_{\rm w},n)$
with $t_{\rm w}=10^{2.5}$ h
\vspace*{2 mm}\\
\noindent
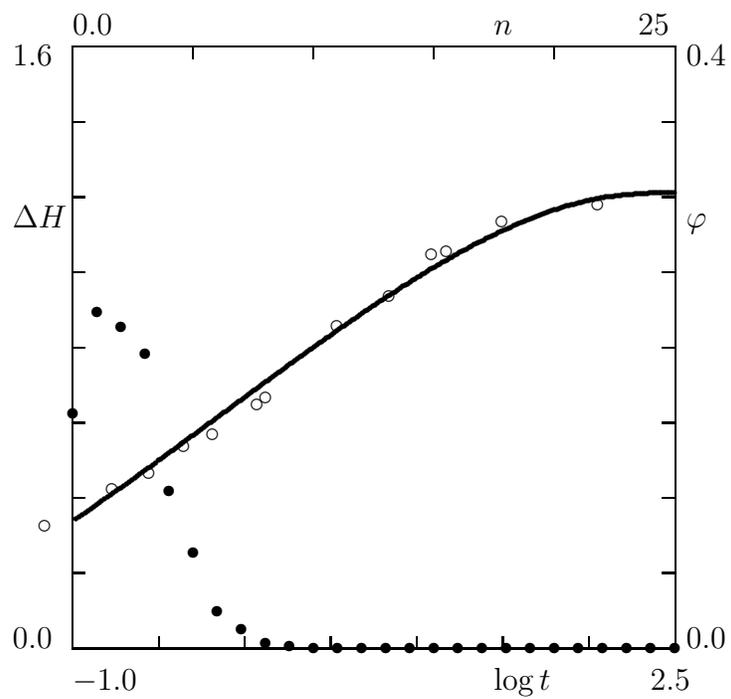
Figure~8: The relaxation enthalpy $\Delta H$ J/g versus time $t$ h
for poly(methyl methacrylate) at $T=382.5$~K
(unfilled circles: experimental data \protect{\cite{CF93}};
solid line: numerical simulation)
and the final distribution of CRRs $\varphi=\varphi(t_{\rm w},n)$
with $t_{\rm w}=10^{2.5}$ h
\vspace*{2 mm}\\
\noindent
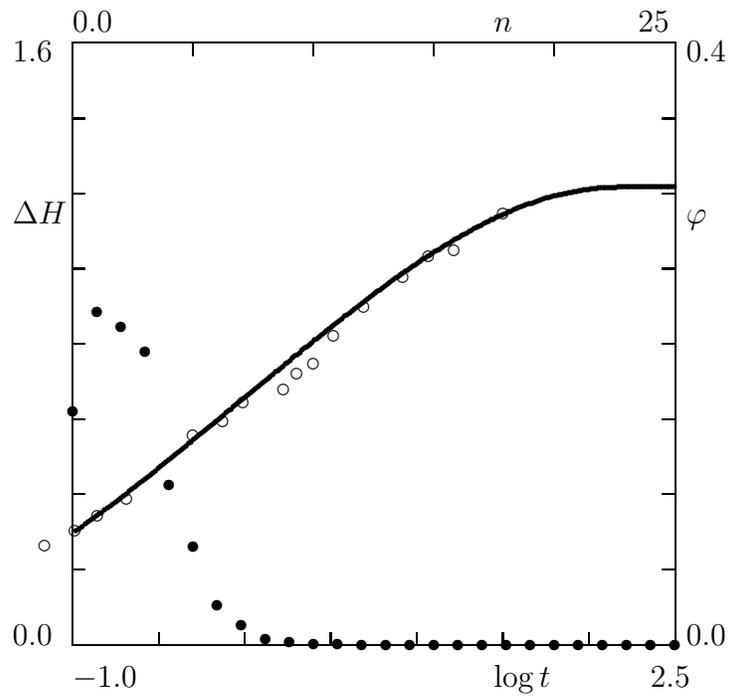
Figure~9: The relaxation enthalpy $\Delta H$ J/g versus time $t$ h
for poly(methyl methacrylate) at $T=385.0$~K
(unfilled circles: experimental data \protect{\cite{CF93}};
solid line: numerical simulation)
and the final distribution of CRRs $\varphi=\varphi(t_{\rm w},n)$
with $t_{w}=10^{2.5}$ h
\vspace*{2 mm}\\
\noindent
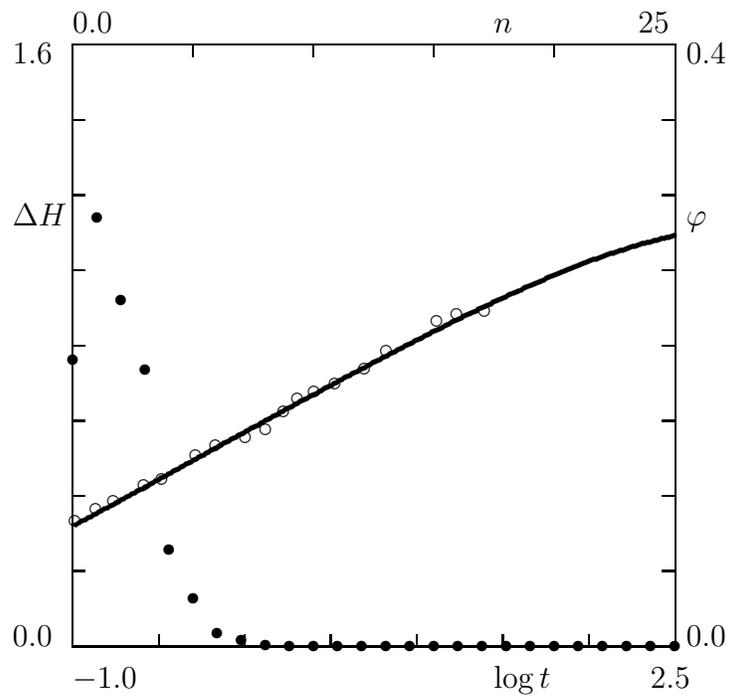
Figure~10: The relaxation enthalpy $\Delta H$ J/g versus time $t$ h
for poly(methyl methacrylate) at $T=387.5$~K
(unfilled circles: experimental data \protect{\cite{CF93}};
solid line: numerical simulation)
and the final distribution of CRRs $\varphi=\varphi(t_{\rm w},n)$
with $t_{\rm w}=10^{2.5}$ h
\vspace*{2 mm}\\
\noindent
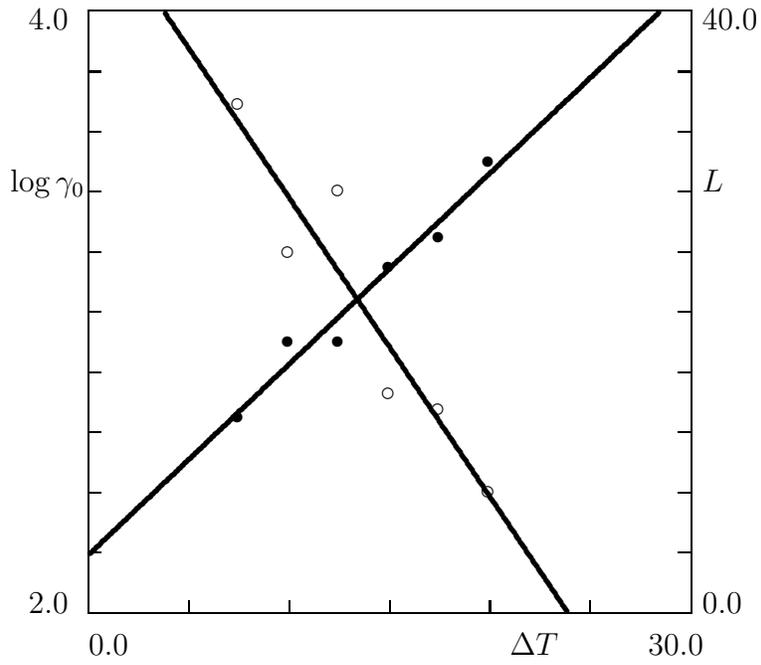
Figure~11: The rate of fragmentation $\gamma_{0}$~h$^{-1}$
(unfilled circles) and the dimensionless parameter $L$
(filled circles) versus the increment
of temperature $\Delta T$~K for poly(methyl methacrylate).
Symbols: treatment of observations \protect{\cite{CF93}}.
Solid lines: approximation of the experimental data by Eqs. (19) and (20)
with $a_{0}=4.3733$, $a_{1}=0.0993$
and $b_{0}=3.8507$ and $b_{1}=1.2686$
\vspace*{2 mm}\\
\noindent
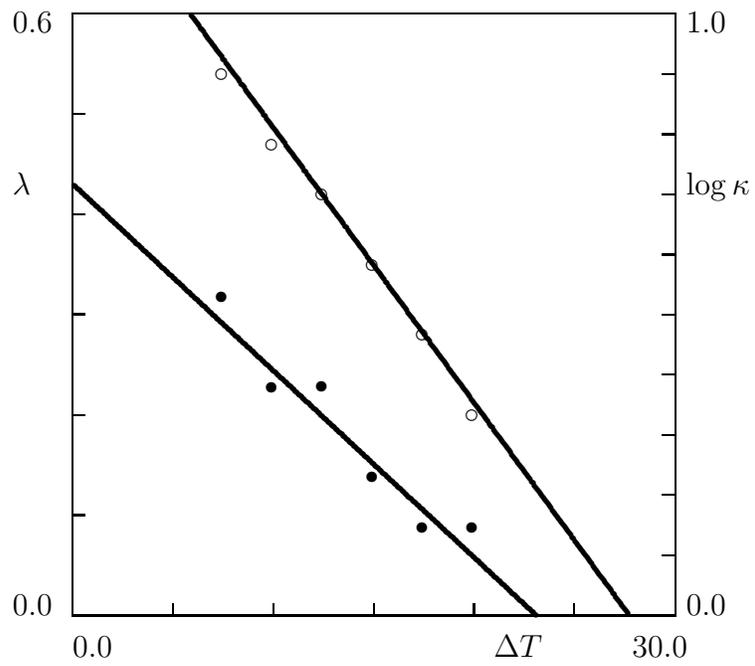
Figure~12: The dimensionless parameters $\lambda$ (unfilled circles)
and $\kappa$ (filled circles) versus the increment of temperature
$\Delta T$~K for poly(methyl methacrylate).
Symbols: treatment of observations \protect{\cite{CF93}}.
Solid lines: approximation of the experimental data by Eq. (21)
with $c_{0}=0.7611$, $c_{1}=0.0274$
and $C_{0}=0.7173$, $C_{1}=0.0310$
\vspace*{2 mm}\\
\noindent
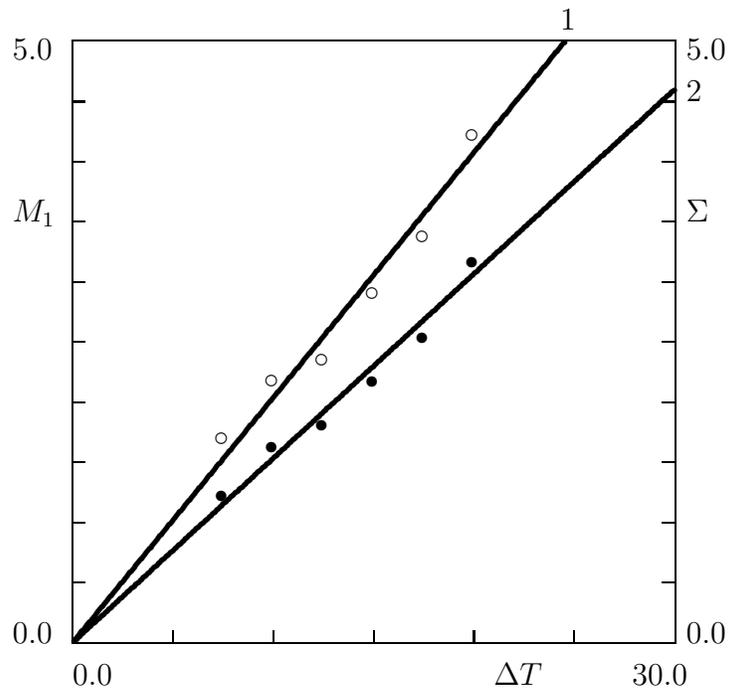
Figure~13: The average number of borders between ECs in a CRR, $M_{1}$,
and its standard deviation, $\Sigma=M_{2}^{\frac{1}{2}}$,
versus the increment of temperature $\Delta T$~K
for poly(methyl methacrylate).
Circles: treatment of observations \protect{\cite{CF93}}.
Solid lines: approximation of the experimental data by Eq. (22)
with $m_{1}=0.2037$ and $m_{2}=0.1533$.
Curve~1: $M_{1}$;
curve~2: $\Sigma$
\vspace*{2 mm}\\
\noindent
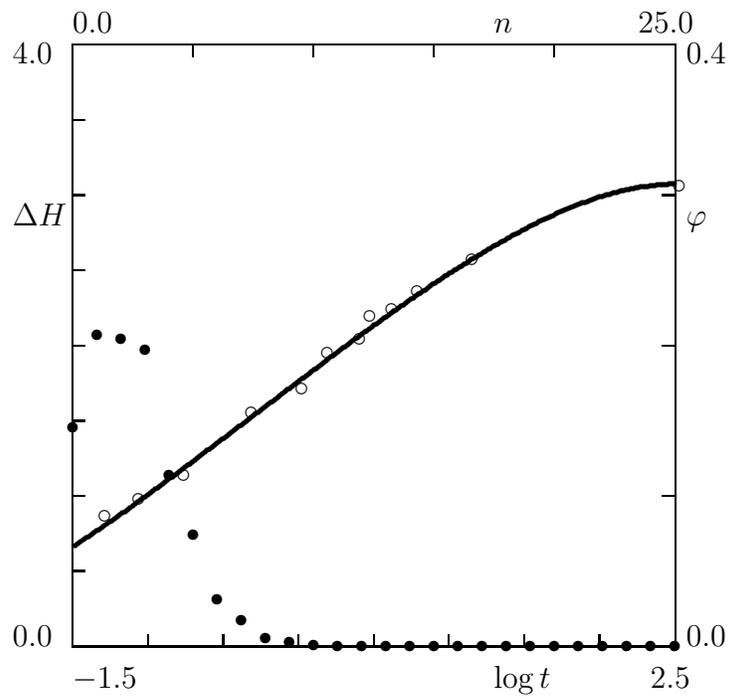
Figure~14: The relaxation enthalpy $\Delta H$ J/g versus time $t$ h
for poly(vinyl acetate) annealed at $T=303$~K.
Unfilled circles: experimental data \protect{\cite{CFH98}}~.
Solid line: numerical simulation with $\Lambda=1.6$ J/g, $\kappa=2.9$,
$L=20.0$, $\lambda=0.45$, $\gamma_{0}=4.74\cdot 10^{2}$ h$^{-1}$.
Filled circles: the distribution of CRRs $\varphi=\varphi(t_{\rm w},n)$
with $t_{\rm w}=10^{2.5}$ h, $M_{1}=2.4209$, $M_{2}=3.2730$
\vspace*{2 mm}\\
\noindent
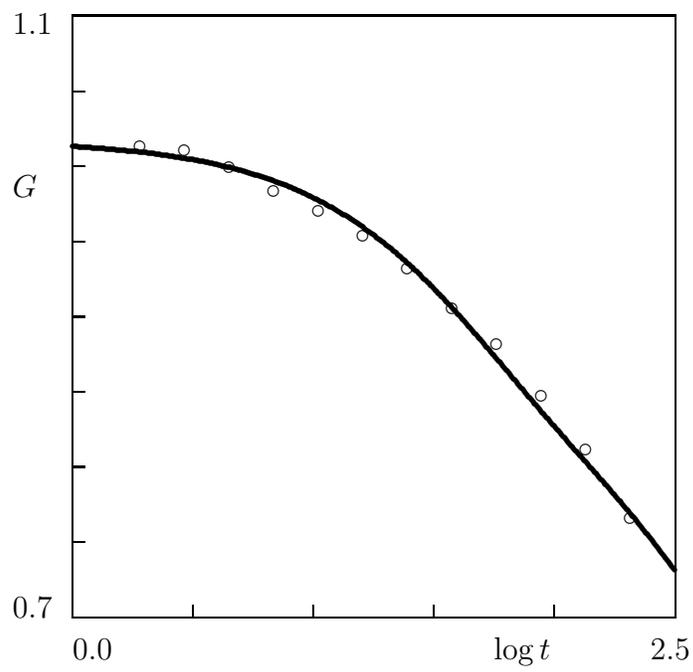
Figure~15: The shear modulus $G$ GPa versus time $t$ s for
poly(vinyl acetate) at $T=303$~K and $t_{\rm w}=16.5$~h.
Circles: experimental data \protect{\cite{CFH98}}.
Solid line: prediction of the model with $\alpha_{0} =2.24$,
$\Gamma_{0}=0.23$~s$^{-1}$ and $G_{0}=1.0176$ GPa
\newpage

\setlength{\unitlength}{0.8 mm}
\begin{figure}[t]
\begin{center}

\end{center}

\caption{}
\end{figure}

\begin{figure}[t]
\begin{center}
%
\end{center}

\caption{}
\end{figure}

\begin{figure}[t]
\begin{center}
%
\end{center}

\caption{}
\end{figure}

\begin{figure}[t]
\begin{center}
%
\end{center}

\caption{}
\end{figure}

\end{document}